\documentclass[lettersize,journal]{IEEEtran}

\usepackage{amsmath,amsfonts}
\usepackage{algorithmic}
\usepackage{array}
\usepackage[caption=false,font=normalsize,labelfont=sf,textfont=sf]{subfig}
\usepackage{textcomp}
\usepackage{stfloats}
\usepackage{url}
\usepackage{verbatim}
\usepackage{graphicx}
\usepackage{booktabs}
\usepackage{xcolor}
\usepackage{cite}
\usepackage{balance}
\usepackage{orcidlink}

\usepackage{cleveref}

\graphicspath{{figures/}{pictures/}{images/}{./}}

\usepackage[table]{xcolor}

\definecolor{clusterreach}{RGB}{225,242,224}   
\definecolor{clusterstatic}{RGB}{255,248,204}  
\definecolor{clusterfine}{RGB}{252,228,214}    

\def\BibTeX{{\rm B\kern-.05em{\sc i\kern-.025em b}\kern-.08em
    T\kern-.1667em\lower.7ex\hbox{E}\kern-.125emX}}

\begin{document}


\title{Altered Visual Hand Feedback in Virtual Reality: An Exploratory Study in Essential Tremor}



%
\author{
Alejandro Pascual-Valdunciel\orcidlink{0000-0002-7984-7516},
Marcos Sacristán-Moraleda\orcidlink{0000-0002-2066-5160},
Verónica Giménez de Béjar\orcidlink{0000-0002-0631-8517},
Julián Benito-León\orcidlink{0000-0002-1769-4809},
Juan Marín-Lahoz\orcidlink{0000-0003-0008-2360},
Jaime Iba\~nez\orcidlink{0000-0001-8439-151X},
Daniel Martin\orcidlink{0000-0002-0073-6398}%
\thanks{Alejandro Pascual-Valdunciel, Marcos Sacrist\'an-Moraleda, and Jaime Ib\'a\~nez are with the Biomedical Signal Interpretation and Computational Simulation (BSICoS) group, Instituto de Investigaci\'on en Ingenier\'ia de Arag\'on (I3A), Universidad de Zaragoza, Zaragoza, Spain.}%
\thanks{Verónica Giménez de Béjar and Julián Benito-León are with the Department of Neurology, 12 de Octubre University Hospital, Madrid, Spain. Julián Benito-León is also with Group of Neurodegenerative Diseases, Hospital Universitario 12 de Octubre Research Institute (imas12), Madrid, Spain; Network Center for Biomedical Research in Neurodegenerative Diseases (CIBERNED), Madrid, Spain; and Department of Medicine, Complutense University of Madrid, Madrid, Spain.}
\thanks{Juan Mar\'in-Lahoz is with the Faculty of Medicine, Universidad de Zaragoza; the Department of Neurology, Hospital Universitario Miguel Servet, Zaragoza; Neuroscience Research group, Instituto de Investigación Sanitaria de Aragón (IISA), Zaragoza, Spain.}%
\thanks{Daniel Martin is with the Graphics and Imaging Lab, Instituto de Investigaci\'on en Ingenier\'ia de Arag\'on (I3A), Universidad de Zaragoza, Zaragoza, Spain.}%
}
\maketitle


\begin{abstract}
    Pathological tremor is the most prevalent movement disorder worldwide, characterized by involuntary muscular oscillations within a frequency of 3 to 12 Hz. Despite extensive research, the underlying mechanisms that generate and modulate tremor are not yet understood, and treatments are usually insufficient. While it is known that visual feedback is an important afferent pathway, little is known about the impact of altering the hand's visual representation on tremor expression. With the development of immersive technologies, VR poses an excellent tool to study how alterations to the representation of the hands affect tremor. In this work, we study this through two complementary experiments. In the first one, we show that a consumer-level hand-tracking system substantially attenuates tremor-band movements without preventing successful task completion. In the second one, with eight patients with essential tremor, we observe task-dependent differences in physical tremor between VR and the real world, with several patients exhibiting lower tremor-band activity in both hand kinematics and muscle oscillations, particularly during reaching tasks. Our findings represent an initial step toward validating the use of VR as an experimental tool for the research and treatment of pathological tremor, while highlighting strong user- and task-dependent effects.
\end{abstract}



%



\section{Introduction}
\label{sec:01_Intro}

\IEEEPARstart{P}{athological} tremor, defined as an involuntary, rhythmic, oscillatory movement of one or more body parts, is a common feature of several neurological disorders and occurs within a frequency range of approximately 3 to 18 Hz. It represents one of the most prevalent movement disorders, affecting millions of people worldwide and imposing substantial personal and societal burdens~\cite{louis2010common}. Essential tremor (ET), the most common form of pathological tremor and typically with a frequency between 4 and 12 Hz, affects approximately 0.9\% of the population, with prevalence increasing to around 5.79\% in individuals over 65 years of age~\cite{louis2021common}. Tremor can heavily impact daily life, since it interferes with everyday activities such as drinking, writing, or manipulating objects. Beyond physical impact, tremor can also have psychological and social consequences, including embarrassment, social withdrawal or anxiety, and therefore an overall reduced quality of life. Furthermore, the mechanisms that generate and modulate tremor are still not completely understood~\cite{montero2025emerging, schaefer2018brain}, and current treatments remain insufficient for many patients~\cite{pascual2022non}.

Although ET involves abnormal oscillatory activity within the central nervous system ~\cite{raethjen2012oscillating, schaefer2018brain}, its expression emerges from the interaction between central oscillators and sensory feedback loops \cite{deuschl2001pathophysiology}. For instance, proprioceptive Ia afferents actively shape tremor output, and recruiting them via sub-threshold electrical stimulation can reduce tremorgenic drive~\cite{gallego2015phase, montero2025emerging}. Beyond proprioception, visual feedback represents another primary afferent pathway, since it is processed by networks that directly overlap with the central circuits implicated in ET ~\cite{liu1999effects, welzel2024interplay}. Indeed, increasing visual feedback gain during motor tasks has been shown to directly amplify tremor amplitude in ET patients~\cite{archer2018widespread}. This suggests an intriguing possibility: changing how users \emph{see} their own hand movement could affect how that movement is \textit{physically expressed}. Studying this in the real world is challenging, if even possible, since the visual motion of our hands is inherently coupled to their physical movement. 

Immersive technologies, such as virtual reality (VR) or mixed reality (MR), however, allow these two signals to be decoupled: a user's physical hand could follow one trajectory while its virtual representation follows another, enabling controlled alterations of visual feedback that would be otherwise impossible in the real world. Previous literature has shown that substantial discrepancies between physical movements and their virtual counterparts can be introduced without users' noticing ~\cite{nilsson201815, Serrano2020_VR-LateralMotion, langbehn2018blink}.

This leaves an open question on whether the involuntary component of hand motion associated with tremor can be altered, and whether such alterations could influence its underlying physical expression. Ideally, the involuntary component of tremor could be visually attenuated while preserving the voluntary component (activity $<$ 4 Hz) required for natural and successful interactions. Given the known influence of visual feedback on tremor expression ~\cite{archer2018widespread,liu1999effects,welzel2024interplay}, we believe this could lead to an alteration of its physical manifestation. 

In this work, we investigate this within a VR setup. We have developed an experimental framework in the Meta Quest 3 (MQ3) headset in which participants perform a set of representative activities of daily living (e.g., simulating eating or holding a smartphone), under matched real-world and virtual conditions. 

Two complementary studies using this setup were conducted to test our hypothesis, including an experiment in patients with ET to evaluate how visually suppressing tremor affects its physical expression. Prior to evaluation in patients, we first validated our framework in a study with healthy participants undergoing functional electrical stimulation (FES)-induced tremor to evaluate (i) whether the selected tasks could be completed seamlessly in VR, (ii) whether they remained feasible when tremor-like movements were induced, and (iii) whether the hand-tracking system effectively attenuated the involuntary high-frequency oscillations while preserving voluntary motor control.

Results from our first experiment showed that the native MQ3 hand-tracking attenuated visually perceived tremor, yet participants could successfully complete all tasks. Our second experiment, involving eight patients with ET, revealed encouraging results, with several activities showing reductions in physical tremor-band power in VR, although its effects are strongly task- and user-dependent.

\section{Related Work}
\label{sec:02_RelatedWork}

\subsection{Perceptual Feedback Alterations}
\label{subsec:02_RelatedWork_Manipulations}

Although immersive systems generally aim to reproduce physical movements faithfully, previous research has shown that substantial discrepancies between real and virtual motion can remain unnoticed and can be exploited to improve user experience and interaction.

A prominent example is redirected walking, where imperceptible manipulations of the virtual camera motion allow users to explore virtual environments larger than the available physical space~\cite{steinicke2008analyses}. Different approaches have investigated curvature~\cite{bolling2019shrinking}, rotation~\cite{schmitz2018you}, and translation gains~\cite{Serrano2020_VR-LateralMotion}, deriving detection thresholds for how much discrepancy can be introduced before users become aware of the manipulation. Other techniques exploit human perceptual mechanisms rather than continuously modifying motion. For example, Langbehn et al.~\cite{langbehn2018blink} leveraged blink-induced suppression to induce changes in position and orientation without users noticing them. Similarly, Sun et al.~\cite{sun2018towards} focused on saccadic suppression to rotate the environment and redirect users. Beyond these short periods of reduced visual sensitivity, previous literature has also studied more general limitations in visual attention and memory. Suma et al.~\cite{suma2011leveraging} exploited change blindness--a phenomenon widely studied in traditional literature~\cite{simons1997change,hollingworth2006scene}--to imperceptibly manipulate a 3D environment and enable the exploration of larger environments. This has motivated subsequent works showing the potential of change blindness in immersive environments~\cite{martin2023study, brubach2023systematic}. All these works suggest that the visual consequences of users' actions can be modified within substantial perceptual limits. 

These principles have also been explored for the visual representation of the user's own body. In particular, hand-redirection techniques have shown that the displayed trajectory of the hand can deviate from its physical counterpart without necessarily being detected~\cite{zenner2019estimating,Hartfill2021,zenner2021blink}. Recent work has further explored how the detectability and effects of these discrepancies depend on the interaction context. For instance, Feick et al.~\cite{feick2024impact} studied hand redirection in relation to avatar embodiment and detectability, and Jang et al.~\cite{jang2024effects} investigated how hand redirection and visual alterations affect targeted interactions. Liao et al.~\cite{liao2025effect} explored the role of hand realism in redirected hand interaction, and Hartfill et al.~\cite{hartfill2024investigating} analyzed the influence of embodiment and realism during hand manipulations. Overall, these works suggest that hand redirection can affect subjective realism and embodiment, while changes in redirection detectability appear to be more limited. Interestingly, beyond general interaction, hand redirection has also been suggested in application-oriented settings such as upper-limb rehabilitation~\cite{xiong2024reach}.

Together, these works show that the visual motion of the hands does not need to reproduce their physical movement faithfully for successful interaction. This suggests that it might be possible to alter visual hand feedback when some components of the movement are unwanted or involuntary.  In this work, we focus specifically on pathological tremor, which is characterized by continuous high-frequency involuntary rhythmic oscillations, and study whether a modified visual expression in VR can affect the underlying physical movement.

\begin{figure*}[!t]
    \centering
    \includegraphics[width=0.98\linewidth]{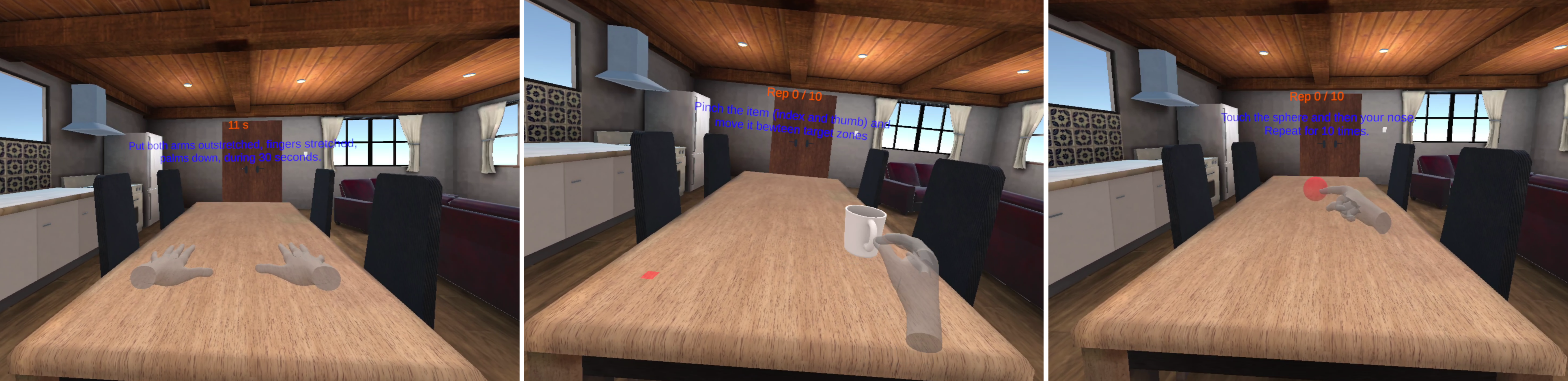}
    \caption{Virtual environment used during our experiment, where participants sat in front of a table within a realistic living room to perform the daily manual tasks studied (Section~\ref{subsec:03_ExperimentalDesign_Tasks}). \textbf{From left to right}: A postural task where participants had to maintain their hands unsupported for 30 seconds, a fine interaction task where participants held and moved a mug between two target zones (semitransparent red zone) ten times, and a reaching task where participants alternated between touching their nose and a predefined point (semitransparent red sphere) intended to make them fully extend their arm, also ten times.}
    \vspace{-10px}
    \label{fig:virtualenvironment}
\end{figure*}

\subsection{Pathological Tremor and Immersive Environments}
\label{subsec:02_RelatedWork_Tremor}
Pathological tremor is generated by abnormal oscillatory activity in central motor networks. In ET, this activity involves the cerebello-thalamo-cortical circuit \cite{schaefer2018brain, raethjen2012oscillating}. However, tremor is not a purely central phenomenon. It emerges from the interaction between central oscillators and the sensory feedback loops that continuously inform the motor system about the movement \cite{deuschl2001pathophysiology, mcauley2000physiological}. Proprioceptive afferents are a key part of these loops. For instance, Ia afferents from muscle spindles signal changes in muscle length and velocity to the spinal cord and the brain, and their input contributes to the tremor observed in ET \cite{gallego2015phase}.
Since sensory feedback shapes tremor expression, manipulating sensory pathways can also modulate tremor. This idea has been tested mainly with electrical stimulation of peripheral nerves, without producing muscle contractions \cite{montero2025emerging}. These inputs act on spinal and supraspinal circuits and can reduce the tremorgenic drive reaching the muscles.

Visual feedback is the other main source of information about our own movements, and it is processed by brain networks that overlap with those involved in ET \cite{archer2018widespread}. However, vision has received much less attention as a target to modulate tremor, partly because the image of the hand cannot be separated from its real movement in the physical world. Most evidence regarding how alterations in visual feedback can affect the manifestation of pathological tremors comes from screen-based tasks, in which participants control a cursor that represents their force or position, and the visual gain (the scaling of the error between the real and the displayed movement) can be changed. In ET, increasing the visual gain in a grip-force task increased 4–12 Hz tremor more than in healthy controls, while force accuracy improved similarly in both groups \cite{archer2018widespread}. The opposite manipulation, reducing the visible tremor, has been tested less. In multiple sclerosis, removing or time-averaging the visual feedback of the moving limb reduced intention tremor \cite{feys2003effect,feys2006effect}. In contrast, removing visual feedback did not change action tremor in Parkinson's disease \cite{liu1999effects}.
VR makes it possible to apply these manipulations to a realistic virtual hand. Cohn et al. showed that a moving-average filter can remove tremor from a virtual hand, but they tested it in a single healthy participant who imitated tremor \cite{cohn2018quantifying}. A preprint reported reduced tremor in nine patients with Parkinson's disease when a stabilized virtual hand was displayed. However, the filtered condition always followed the unfiltered one, so time effects cannot be excluded \cite{cornacchioli2023algorithmic}. Other works are studying how the appearance and movement of virtual hands affect motor cortex activity and functional tremor \cite{kober2026neural,clinicaltrials_NCT06393439}. To our knowledge, no published study has tested whether altering the visual perception of involuntary tremors has an effect on their intensity during simple motor tasks.

\section{Experimental Design and Study Overview}
\label{sec:03_ExperimentalDesign}
Visual feedback has a key role in the way users perceive and control their own movements~\cite{archer2018widespread, liu1999effects, welzel2024interplay}. We hypothesize that reducing the visual manifestation of tremor (i.e., visually removing the high-frequency involuntary component) could effectively reduce pathological tremor. 


The main goal of our work is to investigate whether users exhibit less tremor when their virtual hands appear more stable than their physical hands. To study this question, we have developed a set of simple manual tasks based on everyday activities that are commonly affected by pathological tremor (Section~\ref{subsec:03_ExperimentalDesign_Tasks}). The tasks can be performed both in the real world and in a matched virtual environment (Section~\ref{subsec:03_ExperimentalDesign_Scenarios}), allowing us to compare the resulting movement patterns in both conditions.

We have addressed this through two complementary studies. The first study evaluated the hand-tracking capabilities of the headset and the overall feasibility of the experimental protocol. In particular, it assessed whether users could perform the selected tasks comfortably and complete them successfully within a reasonable amount of time. The second study used the resulting platform to investigate whether stabilized visual feedback affected the exhibition of physical tremor in participants with ET. Both studies shared the same set of tasks, virtual environment, apparatus, data-acquisition pipeline, and general experimental procedure. We therefore describe these common components in this section, while Section~\ref{sec:04_PilotStudy} and~\ref{sec:05_MainExperiment} present any particularities of each study.
\subsection{Tasks}
\label{subsec:03_ExperimentalDesign_Tasks}

We selected six manual tasks inspired by activities of daily living that are commonly affected by pathological tremor. The battery combines clinical examination maneuvers, such as those included in The Essential Tremor Rating Assessment Scale (TETRAS) and the Fahn-Tolosa-Marin Tremor Rating Scale (FTMTRS) \cite{elble2016essential, fahn1993clinical} with tasks inspired by everyday activities and experimental studies. The selected tasks are the following:

\begin{enumerate}

\item \textbf{Postural task.} Participants extended both arms forward, palms downside, and maintained them unsupported and as still as possible for 30 seconds. It is the clinically most extended postural tremor assessment.

\item \textbf{Cup movement with pinch grip.} Participants held a cup using a pinch (index-thumb) grip and moved it repeatedly between two predefined target positions on a table from left to right and back ten times. This task combines fine grasping, object transport, and goal-directed movement.

\item \textbf{Finger-to-nose task.} Participants alternated between touching their nose and reaching forward to touch a predefined target while keeping the index finger extended, repeating the movement ten times. This standard clinical maneuver involves repeated goal-directed movements toward fine targets.

\item \textbf{Cup movement with full-hand grasp.} Participants transported the same cup between the predefined table positions (left and right) but using a full-hand grasp back and forth ten times. Unlike the pinch-grip condition, this task involved using a more stable grasp.

\item \textbf{Spoon feeding task.} Participants performed ten simulated eating movements transporting a spoon from a plate in front of them to their mouth. This task required coordinated hand-arm motion, tool use, and fine motor control.

\item \textbf{Smartphone holding task.} Participants held a mobile phone with both hands in front of them without supporting their arms for 30 seconds. This task combined sustained posture with object holding.

\end{enumerate}

The tasks represented complementary motor contexts in which the relative contribution of voluntary and involuntary movement differed. This allowed us to examine whether the effects of visual alterations affected tremor in a manner contingent on the activity being performed. The order of the activities was fixed (1 to 6), and participants were given a 30-second break between tasks. A video showing these tasks can be found in the supplementary material.

\begin{figure*}
    \centering
    \includegraphics[width=0.96\linewidth]{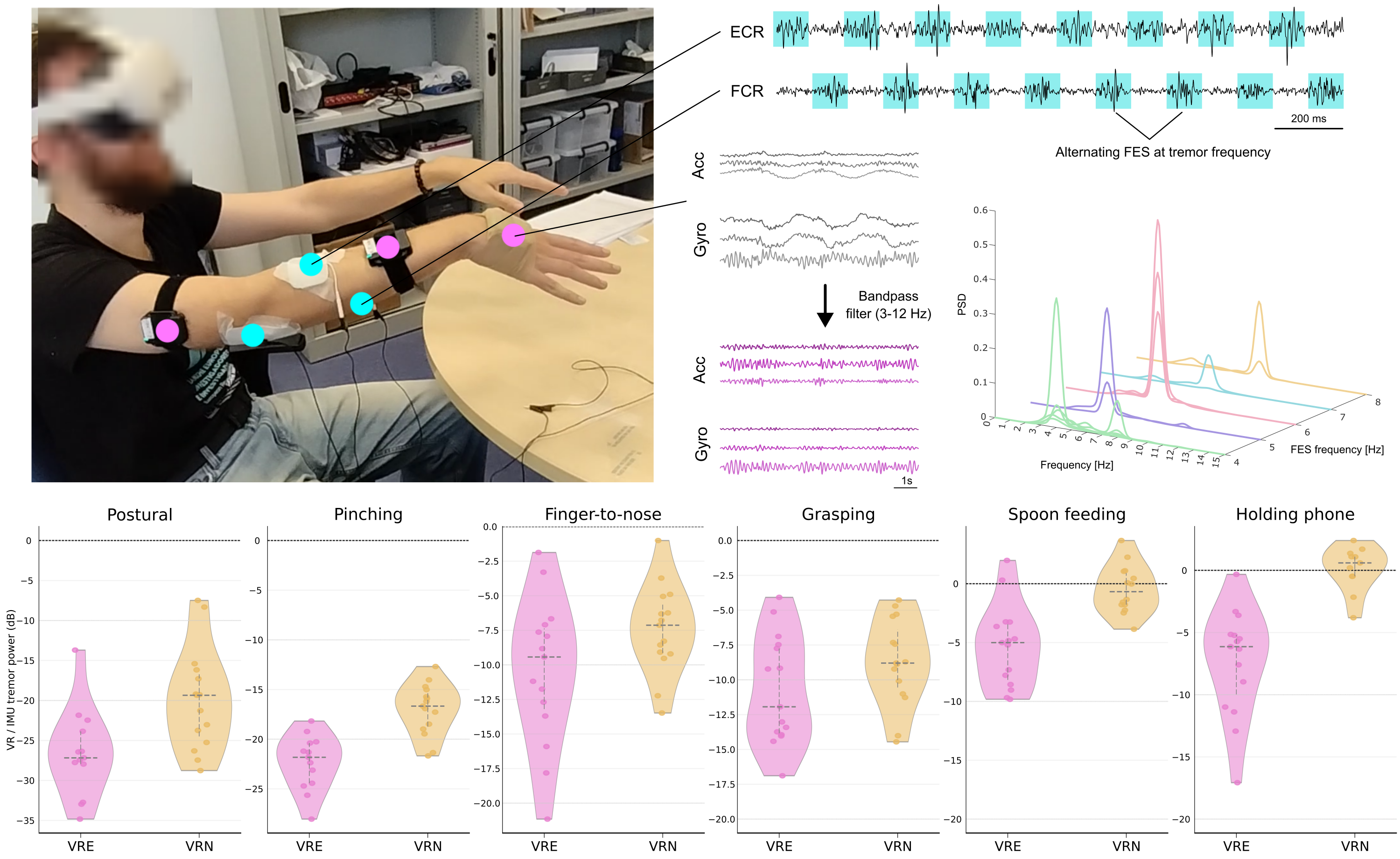}
    \caption{\textbf{Top left}: Experimental setup showing a participant during the study with three x-IMU3 motion sensors (pink circles) and FES electrodes (cyan circles) placed on the extensor carpi radialis (ECR) and flexor carpi radialis (FCR) muscles. 
    \textbf{Top right}: Signal processing pipeline illustrating alternating FES stimulation bursts applied to the ECR and FCR muscles, raw IMU signals (accelerometer and gyroscope), bandpass filtering ($3\text{--}9~\text{Hz}$) to isolate tremor-band activity, and power spectral density (PSD) profiles for every participant across different FES frequencies. 
    \textbf{Bottom}: Visual tremor attenuation introduced by the VR headset, quantified as the power ratio between the visually perceived hand motion in VR and the physical hand motion measured by the IMUs ($\text{VR}/\text{IMU}$, in~$\text{dB}$) across our six functional tasks. Lower (more negative) values indicate greater visual suppression of tremor by the VR display during FES-induced tremor (\textit{VRE}, pink) and standard VR (\textit{VRN}, orange). Violin plots show distribution across participants.}
    \vspace{-10px}
    \label{fig:stimulation_imuvsVR}
\end{figure*}

\subsection{Experimental Scenarios and Stimuli}
\label{subsec:03_ExperimentalDesign_Scenarios}

One of our goals is to determine whether the selected tasks can be reliably reproduced in virtual reality while eliciting hand movements comparable to those observed under real-world conditions. To this end, we conducted our experiments in two different conditions: real world and virtual reality.

For the virtual reality (VR) condition, to elicit realistic behaviors, we designed an environment consisting of a living room (see Figure~\ref{fig:virtualenvironment}) in which the participant was seated in front of a table. The table was empty, and any required objects for the task at hand (e.g., the cup for the cup movement tasks, or the smartphone for the holding task) would spawn at the beginning of the corresponding task and despawn when completed. In this condition, target zones (e.g., where to move the cup) were shown with a semi-transparent red shape, which would locate in the target position of the current movement. A text canvas was always located on the opposite side of the table with the current task's instructions, progress, and timers.

For the real world (RW) condition, we crafted the required objects in paper. While less realistic, this allowed us to perform fair comparisons with the virtual reality condition: real cups' or cutlery's weight could modify the expression of tremor, and thus lead to measurements not directly comparable to the VR condition. In this condition, target zones were either marked in the table with tape (e.g., for the cup movement), or indicated by the experimenter (e.g., the finger-to-nose reaching point), as done in traditional tremor studies ~\cite{fahn1993clinical,elble2016essential}. The experimenter controlled the elapsed time and progress during tasks and rest breaks, periodically informing the participant.

\subsection{Apparatus and Data Logging}
\label{subsec:03_ExperimentalDesign_Apparatus}

We designed the VR condition using Unity 6000.1.9f1. We ran our experiments using a Meta Quest 3 head-mounted display (HMD), which has a spatial resolution of 2064 × 2208 pixels per eye, a field of view (FOV) of 110º × 96º, a temporal resolution of up to 120 Hz, and 25 pixels-per-degree (PPD). The MQ3 headset includes native hand tracking through the Meta Quest SDK. During each task, the system continuously recorded the three-dimensional position and orientation of the HMD, as well as the tracked positions of both hands, including the fingertips and finger bases for all fingers, and the palm. Additionally, the system recorded the grabbing state of each hand and the distance between the thumb and index finger as a measure of pinching.

To capture hand motion independently of the headset, and thus finer tremor information, we used x-IMU3 inertial measurement units (IMUs; x-io Technologies, UK). For the experiment with healthy participants, three IMUs were placed on the dorsal surfaces of the arm, forearm and dominant hand; while for the experiment with tremor patients only an IMU was placed over the dorsum of the hand most affected by tremor. Tri-axial acceleration (full-scale range ±24 g, 16-bit resolution) and angular velocity (full-scale range $\pm2000^\circ$/s, 16-bit resolution) signals were recorded at a sampling rate of 100 Hz . These sensors provided an external high-accurate measurement of hand kinematics that was used to characterize tremor and to compare the physical hand movement against its representation in VR.  

For our second experiment involving participants with pathological tremor, we additionally recorded high-density surface electromyography (HDsEMG) from extensor carpi radialis (ECR) and flexor carpi radialis (FCR) using two 32-channel electrode grids. Both ECR and FCR grids were 8 × 4 electrodes with 8 mm inter-electrode distance. HDsEMG signals were acquired in monopolar configuration (reference over the wrist) with a wireless amplifier (Muovi, OT Bioelettronica, Italy) sampling at 2000 Hz.

\subsection{Common Experimental Procedure}
\label{subsec:03_ExperimentalDesign_Procedure}

All the experiments described in this work followed a common procedure. The study with patients was approved by local ethics committee CEICA (PI26$/$300) and Ethics Committee for Research with Medicines at the Hospital  Universitario 12 de Octubre (24$/$050). The study with healthy volunteers was approved by local ethics committee CEICA (PI23$/$404). All procedures were conducted in accordance with the Declaration of Helsinki. At the beginning of the study, all participants signed the written consent form.

After that, the experimenter attached the set of sensors to monitor tremor (see Section~\ref{subsec:03_ExperimentalDesign_Apparatus}). When the sensor setup was completed, participants were informed of the tasks that they were going to be required to complete (Section~\ref{subsec:03_ExperimentalDesign_Tasks}) in the different scenarios (Section~\ref{subsec:03_ExperimentalDesign_Scenarios}). Then, participants completed both virtual reality and real-world conditions; the order was counterbalanced. In all cases, the presentation order of the six tasks was kept constant.

After completing each condition, they filled out a short version of the NASA-TLX questionnaire~\cite{hart1988development} to assess their perceived physical and mental workload. Besides, for the virtual reality condition, participants filled out a short sickness questionnaire~\cite{kennedy1993simulator} before and after the procedure to assess whether exposure to the virtual environment induced any symptoms of discomfort or cybersickness. Finally, the sensors were removed, and participants completed a short debriefing with the experimenter.

\subsection{Data Processing}
\label{subsec:03_ExperimentalDesign_DataProcessing}
All data were stored and analyzed offline. For the RW condition, kinematic data and task metadata recorded from the MQ3 headset were resampled to 100Hz and synchronized with the data recorded with the IMUs. Here, we focus on the acceleration of the hand, as the final effector of the tasks and where tremor ultimately affects performance. The acceleration was computed from the position of the dominant palm of the hand recorded with the MQ3. Acceleration signals were band-pass filtered (4th order Butterworth, zero-phase) to remove low frequency drifts and high-frequency artifacts and then z-scored to allow comparison across subjects. Then, acceleration signals were filtered in the tremor band (3,9) Hz to isolate the tremor component from the voluntary activity. 
Signal power was characterized in the frequency domain from the power spectral density (PSD) of the hand acceleration signals (pwelch method, 2 s windows, 50\% overlap, 0.1 Hz resolution). Finally, tremor power was estimated as the area of the PSD in the tremor band. Additionally, the maximum value of the PSD and its frequency were also computed. As a complementary measure, angular velocity was also used following a similar analysis pipeline in the experiment with tremor patients. 

To evaluate the effect of FES in healthy participants, we computed the change in the power measured with the IMUs between the \textit{VR-FES} and \textit{VR} conditions. To quantify how much of the physical motion was rendered visually by the headset, attenuation was expressed in decibels as $10\log_{10}(P_{\mathrm{MQ3}}/P_{\mathrm{IMU}})$, where $P_{\mathrm{IMU}}$ and $P_{\mathrm{MQ3}}$ are the tremor-band power measured with the IMUs and its counterpart representation from the Meta Quest 3. For the comparison between RW and VR in patients with ET, we computed within-participant ratios of tremor-band power ($P_{\mathrm{VR}}/P_{\mathrm{RW}}$) for each task, separately for the accelerometer and gyroscope signals. The same procedure was applied to the EMG tremor-band power measurements described below. For the non-time-constrained tasks, we also measured completion time. Discomfort and workload measurements were analyzed using the reported NASA-TLX dimensions and short simulator-sickness questionnaires.

The HDsEMG signals were band-pass filtered between 10 and 500 Hz (4th order Butterworth, zero-phase) and notch filtered at 50 Hz (IIR notch, Q=30, zero-phase). Noisy channels with poor signal-to-noise ratio were discarded based on visual inspection. The EMG envelope of each channel was obtained as the absolute value of its Hilbert transform. Then, the envelope’s mean was removed, and the envelope was band-pass filtered in the tremor band, 3–9 Hz (2nd Butterworth, zero-phase). PSD was then estimated with Welch’s method (pwelch method, 2s windows, 50 \% overlap, 0.1Hz resolution). For each muscle, PSDs from all the valid electrodes were averaged to obtain one PSD per muscle and task. Finally, we computed the tremor power, as the area under the PSD between 3 and 9 Hz.

As all comparisons involved repeated measurements from the same participants (i.e., all participants completed all conditions, randomized order), we used non-parametric repeated-measures tests. Pairwise comparisons were performed using Wilcoxon signed-rank tests, while comparisons across three experimental conditions used Friedman tests followed by paired Wilcoxon tests when deemed necessary. For the study with patients, comparisons of tremor-band power between RW and VR were performed using paired Wilcoxon signed-rank tests on the log-transformed power ratios. Holm correction was applied to p-values to account for multiple comparisons, and Kendall's $W$ is reported as the effect size.

 \section{Validation Study with healthy participants}
\label{sec:04_PilotStudy}

\begin{table}[t]
\centering
\caption{Characterization of the tremor induced by FES (Section~\ref{subsec:04_Pilot_Stimulation}) and its visual attenuation by the Meta Quest 3 (MQ3) hand tracking. $\Delta_{\mathrm{FES}}$ shows the within-participant median change in IMU-derived tremor power between \textit{VR-FES} and \textit{VR} conditions. Attenuation is expressed in dB as \(10\log_{10}(P_{\mathrm{Quest}}/P_{\mathrm{IMU}})\), where \(P_{\mathrm{MQ3}}\) and \(P_{\mathrm{IMU}}\) are the PSD power (i.e., the amount of signal power contained within the tremor-frequency band) of both signals. Negative attenuation values indicate lower tremor power in the visual representation of the hands.} 
\begin{tabular}{lrrr}
\toprule
Task &
$\Delta_{\mathrm{FES}}$ (dB) &
$p_{\mathrm{FES}}$ &
Visual attenuation (dB) \\
\midrule
Postural       & +7.76 & .001 & -19.62 \\
Pinching       & +3.29 & .003 & -22.49 \\
Finger-to-nose & +3.59 & .010 & -19.58 \\
Grasping       & +1.98 & .009 & -18.35 \\
Spoon feeding  & +2.24 & .025 & -15.79 \\
Holding phone  & +7.69 & .001 & -17.38 \\
\bottomrule
\end{tabular}
\label{tab:pilot_psd}
\end{table}

This experiment allowed us to validate the proposed experimental framework in healthy participants under controlled conditions prior to experimentation with tremor patients. Since healthy participants do not exhibit noticeable high-frequency involuntary hand oscillations, we introduced an experimental condition in which we used electrical stimulation to induce a controlled tremor-like movement (Section~\ref{subsec:04_Pilot_Stimulation}) while the participants performed the tasks in the VR scenario. This  condition allowed us to assess whether participants could still complete the proposed tasks even with hand tremor, whether the hand-tracking and sensing systems captured the resulting motion reliably, and whether the experimental pipeline operated correctly under challenging movement conditions.


\subsection{Hand Tremor Stimulation}
\label{subsec:04_Pilot_Stimulation}

To generate controlled tremor-like motion, we used functional electrical stimulation (FES) to induce rhythmic wrist movements in healthy participants (Figure~\ref{fig:stimulation_imuvsVR}, left). Electrical stimulation was applied in an alternating pattern between the wrist flexor and extensor muscles to simulate reciprocal tremor. Adhesive 5 x 5 cm squared electrodes served as anodes over the FCR and ECR muscle belly, and a 5 x 9 cm electrode served as common cathode over the elbow. Stimulation was applied with a carrier frequency of 50~Hz, a pulse width of 400~$\mu$s and muscle-specific intensities (FCR: 9.2 $\pm$ 2.2 mA; ECR: 14.2 $\pm$ 3.7 mA) with a current-constant stimulator (EAST, OT Bioelettronica). Stimulation intensity was adjusted for each participant and muscle to produce visible tremor-like oscillations while remaining below the discomfort threshold. Each participant was stimulated at a single, fixed integer frequency ranging from 4 to 8 Hz, covering frequencies representative of pathological tremor across subjects. This allowed us to generate repeatable oscillatory hand motion simulating pathological tremor while participants performed the same activities used in the other conditions (Section~\ref{subsec:03_ExperimentalDesign_Tasks}).

\subsection{Participants}
\label{subsec:04_Pilot_Participants}

For this pilot experiment, we recruited 15 participants, eight of whom identified as female, six identified as male, and one identified themselves as non-binary. No participant identified as not listed or preferred not to disclose. Their ages ranged from 22 to 40 (mean of 28.4 and STD of 5.38).

\subsection{Procedure}
\label{subsec:04_Pilot_Procedure}

This pilot study adhered to the common procedure described in Section~\ref{subsec:03_ExperimentalDesign_Procedure}, with the only difference that there were three different conditions to evaluate: real world (\textit{RW}), virtual reality (\textit{VR}), and virtual reality with electric stimulation (\textit{VR-FES}) to simulate tremor. The order of the conditions was randomized across participants. 

\begin{table}[t]
\centering
\caption{NASA-TLX median ratings (0-20 scale) across the three conditions (columns). The first row shows the overall score. Higher values indicate greater demand, effort, or frustration, except for performance, where higher means better perceived performance. \textit{VR-FES} yielded the highest workload ratings.}
\small
\resizebox{0.9\columnwidth}{!}{
\begin{tabular}{lrrrrr}
\toprule
Measure & \textit{RW} & \textit{VR} & \textit{VR-FES} & $W$ & $p$ \\
\midrule
Overall workload & 1.67 & 1.83 & 4.33  & .76 & $<.001$ \\
\midrule
Mental demand    & 2    & 2    & 3     & .56 & $<.001$ \\
Physical demand  & 2    & 3    & 5     & .59 & $<.001$ \\
Temporal demand  & 0    & 0    & 2     & .23 & .032 \\
Performance      & 20   & 19   & 17    & .50 & .001 \\
Effort           & 3    & 3    & 8     & .84 & $<.001$ \\
Frustration      & 0    & 1    & 3     & .64 & $<.001$ \\
\bottomrule
\end{tabular}
}
\label{tab:pilot_tlx}
\end{table}

\subsection{Results}
\label{subsec:04_Pilot_Results}

\paragraph{Induced tremor and hand-tracking response} 

First, to characterize how the Meta Quest 3 hand-tracking system represents tremor-like motion (i.e., high-frequency involuntary oscillations), we verified that FES (Section~\ref{subsec:04_Pilot_Stimulation}) generated a measurable increase in physical tremor-band motion. IMU-derived tremor power was significantly higher in the \textit{VR-FES} condition than in \textit{VR} for all six tasks (paired Wilcoxon signed-rank tests; all Holm-corrected $p\leq .025$; see Table~\ref{tab:pilot_psd} and Figure~\ref{fig:stimulation_imuvsVR}, bottom). The within-participant median increase ranged from 1.98~dB for grasping (the lowest) to 7.76~dB for the postural task (the highest), confirming that stimulation produced the desired tremor-like motion across all the studied tasks. We show additional visualizations in the supplementary material.

We next examined how this physical motion was visually rendered by the headset. Within \textit{VR-FES}, the MQ3 visual representation of the hands had a significantly lower tremor-band power than the corresponding IMU measurements for every task (all Holm-corrected $p<.001$). The magnitude of this attenuation was nonetheless task dependent, with median reductions ranging from 22.49~dB for pinching to 15.79~dB for spoon feeding. These results are coherent with the different magnitudes of FES-induced mechanical tremor induced across motor tasks.

Importantly, although FES significantly changed the physical hand motion, this increase was mostly absent from the motion visually represented by the headset: power did not significantly differ between \textit{VR-FES} and \textit{VR} for any of the six tasks after correction for multiple comparisons within the MQ3 hand-tracking system. Together, these results show that the native Meta Quest 3 hand-tracking pipeline substantially attenuated the tremor-like high-frequency motion introduced by FES.

\paragraph{Task feasibility and completion time} 

All participants successfully completed the experimental protocol under the three conditions (\textit{RW}, \textit{VR}, and \textit{VR-FES}). We have analyzed whether reproducing the real tasks in VR or introducing tremor-like motion affected task performance. We analyzed this for four of the implemented tasks: pinching, finger-to-nose, grasping, and spoon feeding. The postural and smartphone holding tasks are not included in this analysis because their duration is fixed to 30 seconds by experimental design.


\begin{table*}[!t]
\centering
\small
\caption{Comparison of tremor power between virtual reality (VR) and the real world (RW) during our second experiment (N = eight patients). \textit{Band power} represents the total oscillatory power integrated over the tremor frequency band, while \textit{peak power} corresponds to the maximum value of the power spectral density (PSD), characterizing the strength of the dominant oscillatory component. Values are median within-participant VR/RW ratios (i.e., ratios lower than 1 (boldfaced) mean less tremor in VR), with Q1-Q3 in brackets. $n_\downarrow$ indicates the number of participants exhibiting lower power in VR. Rows are shaded according to shared trends found based on the type of task (see experiment discussion): reaching tasks (green), static maintenance tasks (yellow), and fine interaction tasks (red).} 

\label{tab:patient-power}
\resizebox{\textwidth}{!}{%
\begin{tabular}{lcccccccc}
\toprule
& \multicolumn{4}{c}{Band Power}
& \multicolumn{4}{c}{Peak Power} \\
\cmidrule(lr){2-5}\cmidrule(lr){6-9}
& \multicolumn{2}{c}{Accelerometer}
& \multicolumn{2}{c}{Gyroscope}
& \multicolumn{2}{c}{Accelerometer}
& \multicolumn{2}{c}{Gyroscope} \\
\cmidrule(lr){2-3}\cmidrule(lr){4-5}
\cmidrule(lr){6-7}\cmidrule(lr){8-9}
Task
& VR/RW [Q1--Q3] & $n_\downarrow$
& VR/RW [Q1--Q3] & $n_\downarrow$
& VR/RW [Q1--Q3] & $n_\downarrow$
& VR/RW [Q1--Q3] & $n_\downarrow$ \\
\midrule
\rowcolor{clusterstatic}
Postural
& 1.445 [0.581--1.737] & 3/7
& 1.158 [0.432--2.163] & 3/7
& \textbf{0.784} [0.499--1.914] & 4/7
& \textbf{0.975} [0.412--2.410] & 4/7 \\
\rowcolor{clusterfine}
Pinching
& 1.669 [1.088--2.148] & 1/8
& 1.219 [0.738--2.151] & 4/8
& 1.960 [1.680--2.057] & 1/8
& 1.222 [0.795--1.834] & 3/8 \\
\rowcolor{clusterreach}
Finger to nose
& \textbf{0.752} [0.386--1.083] & 6/8
& \textbf{0.528} [0.286--0.879] & 7/8
& \textbf{0.881} [0.527--0.978] & 7/8
& \textbf{0.475} [0.387--0.760] & 6/8 \\
\rowcolor{clusterfine}
Grasping
& 1.753 [1.255--2.134] & 2/8
& 1.170 [0.787--2.339] & 4/8
& 1.640 [1.300--2.094] & 1/8
& 1.238 [0.516--2.191] & 4/8 \\
\rowcolor{clusterreach}
Spoon feeding
& \textbf{0.757} [0.322--1.399] & 6/8
& \textbf{0.427} [0.199--0.650] & 7/8
& \textbf{0.571} [0.403--1.167] & 6/8
& \textbf{0.410} [0.191--0.776] & 7/8 \\
\rowcolor{clusterstatic}
Holding phone
& 1.073 [0.166--1.896] & 4/8
& \textbf{0.504} [0.123--3.187] & 5/8
& \textbf{0.944} [0.209--2.336] & 4/8
& \textbf{0.447} [0.210--4.338] & 5/8 \\
\bottomrule
\end{tabular}%
}
\end{table*}

Completion time significantly differed across conditions for the pinching ($\chi^2(2)=20.04$, $p_{\mathrm{Holm}}<.001$, $W=.77$) and grasping tasks ($\chi^2(2)=16.98$, $p_{\mathrm{Holm}}<.001$, $W=.65$). For pinching, median completion time increased from 44.64 s. in \textit{RW} to 73.30 s. in \textit{VR} and 82.55 s. in \textit{VR-FES}. Similarly, for grasping, median completion time increased from 44.18 s. in \textit{RW} to 82.54 s. in \textit{VR} and 78.37 s. in \textit{VR-FES}. Pairwise comparisons showed that both VR conditions were significantly slower than \textit{RW} for both tasks (all Holm-corrected $p<.001$), while no significant differences were found between \textit{VR} and \textit{VR-FES} (pinching: $p=.424$; grasping: $p=.910$). This suggests that, although participants needed more time to complete the tasks in VR, that did not come from the stimulation but instead from the change in medium. We hypothesize that low familiarity with virtual hand-tracking and small tracking errors in finer movements (such as pinching) may have caused this. No significant overall effect of condition was found for finger-to-nose ($\chi^2(2)=5.06$, $p_{\mathrm{Holm}}=.092$, $W=.19$) or spoon feeding ($\chi^2(2)=6.16$, $p_{\mathrm{Holm}}=.092$, $W=.24$). We show a visualization of these results in the supplementary material.

Overall, these results indicate that reproducing some daily manual tasks in a VR medium slightly increased completion time relative to the real-world condition, but introducing FES-induced tremor did not produce an additional measurable cost, thus suggesting that tasks should be feasible even for patients with tremor.

\paragraph{Participant workload and discomfort} 

We have also evaluated whether exposure to the virtual environment, and in particular the addition of FES, produced any discomfort or increased perceived workload. Participants completed a short simulator sickness questionnaire (SSQ~\cite{kennedy1993simulator}) before and after both the \textit{VR} and \textit{VR-FES} conditions. Symptoms were overall very low: the median score was zero both before and after each condition. We found no significant pre-to-post change for either \textit{VR} ($p_{\mathrm{Holm}}=.952$) or \textit{VR-FES} ($p_{\mathrm{Holm}}=.952$), and the magnitude of the change did not differ between the two conditions ($p_{\mathrm{Holm}}=1.000$) either. Overall, neither VR exposure nor the addition of FES produced measurable increases in sickness or discomfort.

We also assessed perceived workload using NASA-TLX~\cite{hart1988development}. Overall workload differed significantly across conditions ($\chi^2(2)=22.91$, $p<.001$, $W=.76$). To maintain consistency with the individual NASA-TLX dimensions, all scores are reported on a 0--20 scale. Median overall workload was lowest in RW (1.67), slightly higher in VR (1.83), and significantly highest in VR-FES (4.33). Pairwise comparisons showed significant differences between all conditions. Similar patterns were observed across the individual NASA-TLX dimensions (Table~\ref{tab:pilot_tlx}), with the largest effects observed for effort ($W=.84$), frustration ($W=.64$), and physical demand ($W=.59$). Overall, VR-FES was perceived as more demanding than the other conditions. This additional workload highlights the burden introduced by tremor's involuntary motion during everyday manual tasks, clearly motivating the need to investigate if altering its visual representation can reduce its impact. We explore this in the following section.

\section{Pilot Study in Essential Tremor Patients}
\label{sec:05_MainExperiment}

Our first study (Section~\ref{sec:04_PilotStudy}) established that our experimental framework could be used to successfully reproduce daily manual tasks in virtual reality and, importantly, showed that the virtual representation of the hands attenuated high-frequency involuntary motion (i.e., movements above approximately 3 Hz) while preserving the lower-frequency voluntary component required to complete the tasks.

Building on these results, we conducted a second study with participants clinically affected by ET. Participants completed the same set of activities described in Section~\ref{subsec:03_ExperimentalDesign_Tasks} under matched real-world and virtual-reality conditions, following the common experimental procedure described in Section~\ref{subsec:03_ExperimentalDesign_Procedure}. The goal of this experiment was twofold: first, to evaluate our framework within an ecologically valid context, with real patients who are, generally, not used to immersive technologies; and second, to investigate whether reducing the visual expression of tremor could influence its underlying physical manifestation by comparing the physical expression of tremor across the real (\textit{RW}) and virtual (\textit{VR}) conditions.

\begin{figure*}
    \centering
    \includegraphics[width=1.0\linewidth]{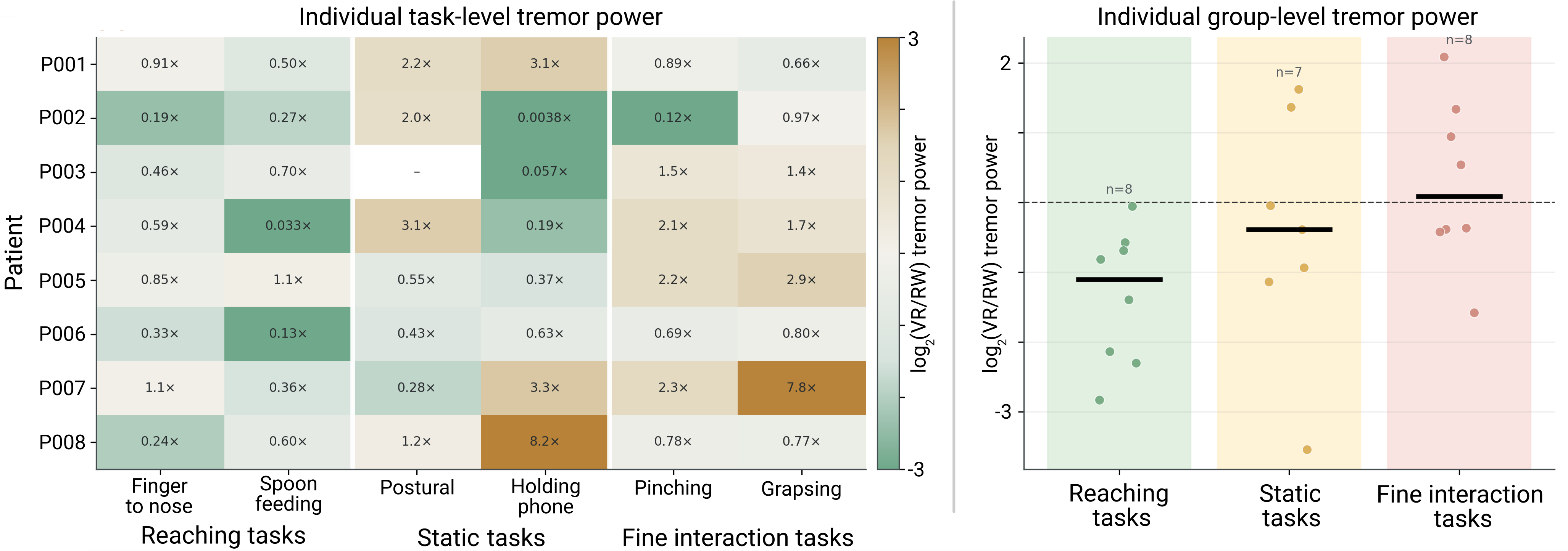}
    \caption{Task-dependent changes in tremor-band power measured with the IMU gyroscope in tremor patients. \textbf{Left}: Per-participant (y-axis) VR/RW tremor-band power ratios across six tasks (x-axis). Values below 1 indicate lower measured tremor power in VR. Colors encode the log$_2$-transformed ratios. Tasks are organized into three categories: reaching (finger-to-nose and spoon feeding), static maintenance (postural and holding phone), and fine interaction (pinching and grasping). The missing value corresponds to corrupted data for P003 in the postural task. \textbf{Right}: Responses aggregated within each of the three task groups aforementioned, computed as the geometric mean of the VR/RW ratios of its two corresponding tasks, and displayed also on a log$_2$ scale. Each point represents one patient, black lines are group medians, and the dashed line indicates equal power between conditions (i.e., same tremor in VR and RW).}
    \vspace{-10px}
    \label{fig:exp2_results}
\end{figure*}

\subsection{Participants}

Eight patients with essential tremor participated in the study (four identified as male and four as female), aged between 58 and 85 years (M = 68.8, SD = 10.6). Five participants were recruited at Hospital Universitario Miguel Servet (Zaragoza, Spain), and three at Hospital Universitario 12 de Octubre (Madrid, Spain). The sample included patients with heterogeneous tremor manifestations and varying levels of severity, as indicated by their clinical observations.

\subsection{Results}

\paragraph{Kinematic outcomes (IMUs)}
We have first evaluated physical tremor using the within-participant ratio between the power of the gyroscope and accelerometer signals within the tremor-frequency band in the VR and RW conditions. Table~\ref{tab:patient-power} shows the median ratios and interquartile ranges across participants, together with the number of participants that showed lower power in VR. Although no individual task showed a statistically significant difference between VR and RW after correction (all $p_{\mathrm{Holm}}\geq .05$), interesting task-dependent trends can be observed, with some activities showing relatively consistent reductions in tremor-band power across participants within the VR condition, while others exhibited more heterogeneous responses. We delve into these patterns below.

\paragraph{Task-level changes in tremor-band power}
As aforementioned, the direction and magnitude of the VR/RW differences varied across activities (Figure~\ref{fig:exp2_results}, left, shows per-participant gyroscope details, and the supplementary material reviews accelerometer data). The most consistent reductions in gyroscope tremor-band power were observed for finger-to-nose movements (median VR/RW ratio $0.528$, with lower power in $7/8$ participants) and spoon feeding ($0.427$, $7/8$ participants). These correspond to median ratios approximately $47\%$ and $57\%$ below the RW tremor, respectively. Nevertheless, the individual comparisons did not remain significant after correction ($p_{\mathrm{Holm}}=.094$). 


The postural and phone-holding tasks showed slightly more heterogeneous responses. For postural holding, the median gyroscope ratio was $1.158$, with $3/7$ participants exhibiting lower tremor-band power in VR. In contrast, holding the phone yielded a median ratio below one ($0.504$), with $5/8$ participants showing reductions. 
Pinching and grasping exhibited median gyroscope ratios of $1.219$ and $1.170$, respectively, and lower power in VR for only $4/8$ participants in each task. 
Details of the accelerometer data can be found in the supplementary material.

\paragraph{Task-dependent differences}
The differences observed above suggest that the effect of VR on physical tremor may depend on the motor demands of each activity. To better characterize these patterns, we have grouped the six tasks \emph{post hoc} into three categories: reaching tasks (finger-to-nose and spoon feeding), static maintenance (postural and holding phone), and fine object interaction (pinching and grasping interactions). Figure~\ref{fig:exp2_results}, right, shows these groups by aggregating the two task-level VR/RW ratios using a geometric mean for the gyroscope data. Accelerometer details are in the supplementary material. The median gyroscope ratios per group were $0.475$ for reaching, $0.763$ for static maintenance, and $1.115$ for fine interaction, with lower power in VR for $8/8$, $5/7$, and $4/8$ participants, respectively. We have also compared the log-transformed VR/RW band-power ratios across the four individual movement tasks (pinching, finger-to-nose, grasping, and spoon feeding). Significant differences were found for 
the gyroscope data ($\chi^2(3)=11.10$, $p=.011$, $W=.463$).

\begin{figure}
    \centering
    \includegraphics[width=0.95\linewidth]{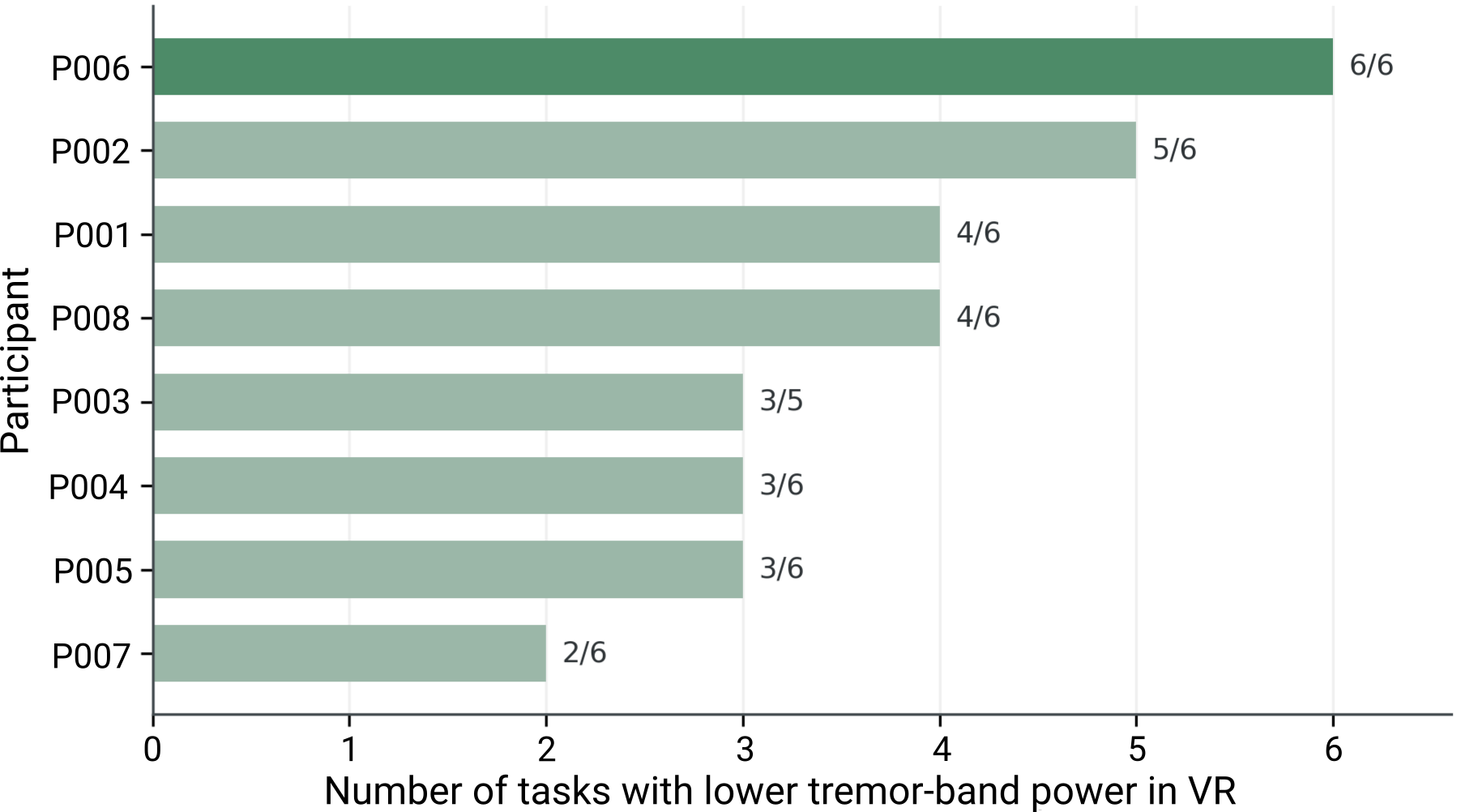}
    \caption{Number of tasks in which each patient (y-axis) exhibited lower gyroscope tremor-band power in VR than in the real world (VR/RW $< 1$, x-axis). Our results highlight significant individual variability in both the direction and consistency of tremor changes within VR, and reveal that some patients exhibited remarkably consistent reductions in tremor-band power across tasks when using VR, showing potential as a tool for further studies.}
    \vspace{-10px}
    \label{fig:exp2_patient}
\end{figure}

\paragraph{Inter-individual variability}
Tremor power not only depended heavily on the task, as aforementioned, but also on the patient. Figure~\ref{fig:exp2_patient} provides a complementary view of the responses, showing, for each patient, the number of tasks in which gyroscope tremor-band power was lower in VR. The counts varied considerably between individuals: for instance, P006 exhibited lower power in all six tasks, P002 in five, while P007 showed reductions in only two. This variability, together with the differences in magnitude depicted in Figure~\ref{fig:exp2_results}, highlights that changes associated with the VR condition were neither uniform across tasks nor consistent across participants.


\paragraph{Task-completion time}
We analyzed the four activities that are not time-constrained. All eight participants took longer in VR than in RW for every one of these tasks. The median within-participant VR/RW duration ratios were $2.37$ for pinching, $1.50$ for finger-to-nose, $2.30$ for grasping, and $2.44$ for spoon feeding. Every comparison remained significant after Holm correction across the four tasks (paired Wilcoxon tests; all $p_{\mathrm{Holm}}=.031$). These results, further shown in the supplementary material, highlight a substantial time cost when performing the current set of movements in VR in the current setting. However, the size of this difference should be interpreted with caution. Participants did not receive constrained instructions on task speed, and many of them performed the RW tasks quickly, which likely increased the VR/RW ratios. Additionally, lack of experience with the VR medium or tracking errors could contribute to the observed differences.

\begin{figure}
    \centering
    \includegraphics[width=0.9\linewidth]{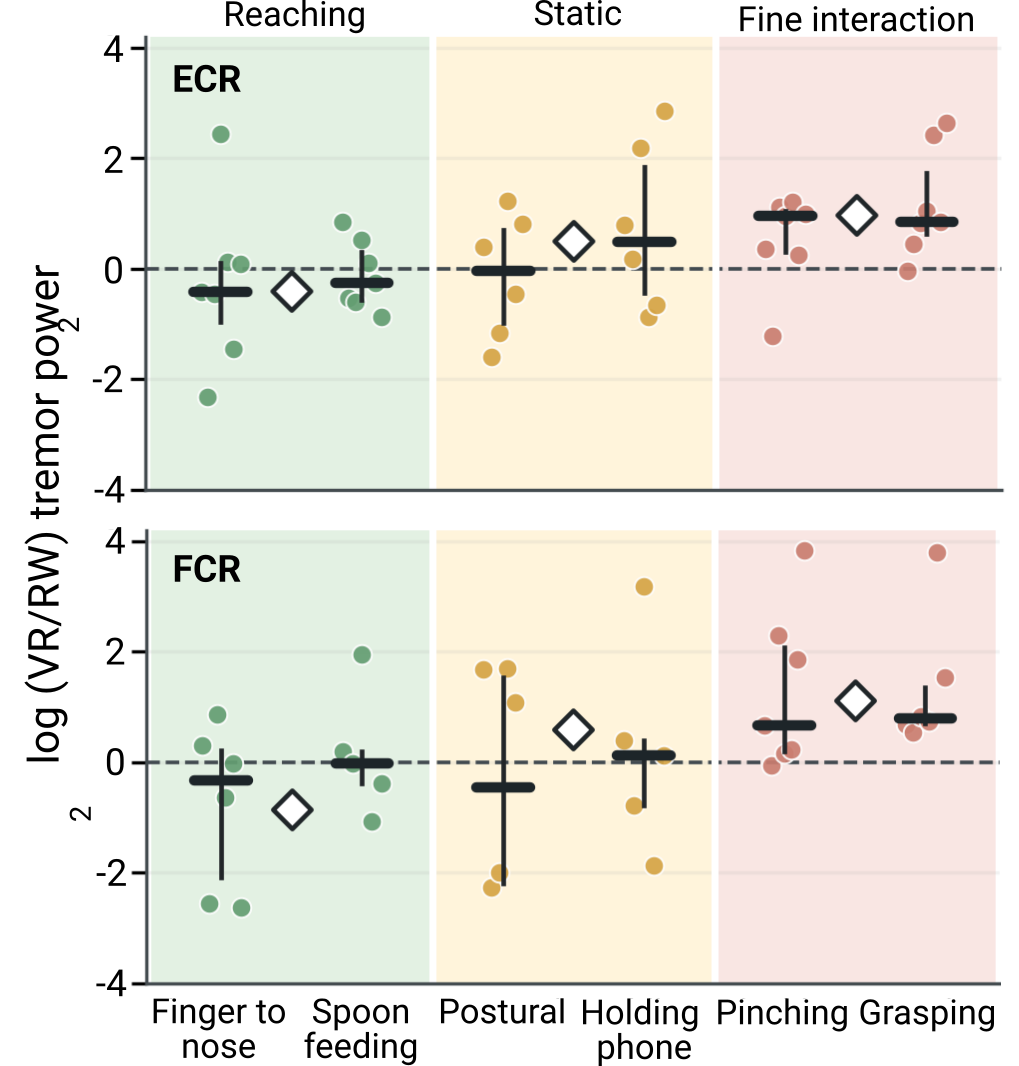}
    \caption{Changes in tremor-band muscle activity between virtual reality (VR) and real-world (RW) conditions in tremor patients, measured using electromyography (EMG) from the extensor carpi radialis (ECR, \textbf{top}) and flexor carpi radialis (FCR, \textbf{bottom}). Each point represents a participant's VR/RW tremor-power ratio for one task (log$_2$-transformed), with negative values indicating lower power in VR. Tasks are grouped as in previous figures. Black bars are medians, and hollow diamonds represent the median of the within-participant geometric mean ratios across the two tasks in each category. Our results support insights from kinematic data (Figure~\ref{fig:exp2_results}).}
    \vspace{-10px}
    \label{fig:exp2_emg}
\end{figure}

\paragraph{Neuromuscular activity (EMG)} We complemented our previous results on kinematics with an additional analysis of the tremor-band power computed from the EMG envelopes of the wrist extensor and flexor muscles. The two measures are related but not equivalent, since the activity of other muscles and the biomechanics shape the observed tremor, while EMG reflects the tremorgenic neural drive that reaches the muscles. Paired RW-VR measurements were available for seven of the patients. Following similar trends to our kinematic results, we observed strong task-dependent effects (Figure~\ref{fig:exp2_emg}, see also supplementary material). For finger-to-nose and spoon feeding, median VR/RW EMG power ratios were $0.748$ and
$0.839$ in ECR, and $0.812$ and $0.983$ in FCR, respectively. In contrast, object-interaction tasks showed the largest increase in EMG tremor-band power in VR, with ratios of $1.933$ and $1.804$ in ECR for pinching and grasping, and $1.584$ and $1.725$ in FCR, respectively. When exploring the aforementioned task grouping, we also found the lowest median ratios in reaching (ECR: $0.762$; FCR: $0.551$) and the highest in fine interaction (ECR: $1.957$; FCR: $2.279$), showing compatible trends again. Interestingly, when comparing patterns across patients, ECR tremor-band power was strongly correlated with the gyroscope measurmements (Spearman's $\rho=.964$, p=$.0028$. These findings indicate that the changes in tremor kinematics observed in VR are accompanied by changes in the tremorgenic activity projected to the task-engaged muscles.

\paragraph{Subjective experience and acceptability}
We have assessed perceived workload using NASA-TLX (see supplementary material). Overall workload remained relatively low in both conditions, with median scores of 2.00 in RW and 2.83 in VR (0-20 scale). No significant difference was found between conditions (paired Wilcoxon test, $p=.500$). Four participants reported higher workload in VR, three reported lower workload, and one showed no difference. Interestingly, perceived performance tended to be lower in VR, which is consistent with the longer time for task completion. However, no difference in any TLX dimension was statistically significant. Overall, our results show that neither the required additional time in VR nor the medium itself was accompanied by increases in perceived workload, thus suggesting that VR may be a tool well tolerated for studying pathological tremor.

\section{Discussion}
\label{sec:06_Discussion}
This work provides several insights into the use of VR to modify physical tremor expression in immersive environments. First, in healthy participants, the native hand tracking of the Meta Quest 3 (MQ3) system attenuated FES-induced tremor-band motion of the virtual hands by approximately 16 to 22 dB, while participants could still complete all tasks. Second, in eight patients with ET, tremor-band power was lower in VR than in the real world (RW) during reaching tasks, with median reductions of $47\%$ (finger-to-nose) and $57\%$ (spoon feeding) and lower power in seven of the eight patients. In contrast, static and fine-interaction tasks did not show consistent reductions. Although individual comparisons did not survive correction for multiple testing, these results suggest that altered visual feedback of the hands may modulate physical tremor, and that this effect depends on both the task and the patient. Overall, this highlights a promising potential of VR technologies for the study and, potentially, treatment of pathological tremor.


Using FES to induce controlled tremor in healthy participants, we found that most of the increase in tremor-band power was absent from the visual representation of the hands. Previous studies have characterized the spatial accuracy of consumer hand tracking~\cite{abdlkarim2024methodological}, but not how it represents involuntary high-frequency motion. The strength of this attenuation was task-dependent: the largest median reduction was observed for pinching (22.49 dB), whereas the smallest reduction was observed for spoon feeding (15.79 dB). This suggests that the extent to which involuntary motion disappears from the visual representation of the hands depends on the voluntary movement being performed, which is particularly relevant when studying motor behavior.

Our results also support the feasibility of reproducing tasks clinically relevant for studying tremor in virtual environments. All our participants could successfully complete all tasks in the three experimental conditions. Some tasks required more time in VR than in the real world (e.g., pinching or grasping), but introducing FES-induced tremor did not result in an additional measurable cost in time completion. These results suggest that the additional time required for some tasks in the virtual conditions was primarily related to difficulties in fine object interaction through virtual hand tracking, rather than the induced tremor itself. We hypothesize that small tracking inaccuracies, unfamiliarity with virtual hand tracking systems, and the lack of physical contact with the virtual objects (i.e., realistic haptic responses) may all contribute to this difference.

During our experiments, we also collected subjective measurements of discomfort and workload. We found that exposure to our virtual environment did not lead to any measurable increase in sickness or discomfort, supporting the suitability of our protocol. Perceived workload, however, increased significantly in the \textit{VR-FES} condition, particularly for effort, frustration, and physical demand. Although we cannot attribute this increase exclusively to tremor (i.e., the electrical stimulation itself may have contributed to perceived effort and discomfort), we believe that it reflects the fact that tremor induces additional motor challenges. 

Our second study extended our findings to patients with ET, allowing us to investigate whether reduced visual expression of tremor could lead to changes in its physical manifestation. Our results revealed a heterogeneous pattern. On the one hand, some activities showed promising trends. In particular, finger-to-nose and spoon-feeding movements (i.e., reaching movements) exhibited lower tremor-band power in seven of the eight patients, while tasks requiring maintaining certain positions (i.e., postural and holding a phone) did not achieve consistently lower tremor across patients. Although several of the comparisons did not survive statistical tests, these findings suggest that the relationship between altered visual hand feedback and expressed physical tremor may depend on the motor demands of the activity being performed. The EMG results provide complementary evidence that the changes observed in hand kinematics may be accompanied by changes in oscillatory muscle activity, rather than being limited to only the representation of the hands.


One possible explanation for our results is that the relative contribution of visual and proprioceptive feedback differs across these motor contexts. Reaching movements rely on continuous visual feedback for online corrections, especially when approaching the target~\cite{sober2005flexible}. In ET, tremor increases when the visual feedback of movement errors is amplified~\cite{welzel2024interplay, archer2018widespread}. A virtual hand without visible tremor may therefore reduce the visually-driven corrections contributing to tremor oscillations during reaching. In contrast, maintaining a static posture without an explicit visual target relies mostly on proprioception, which may explain the smaller effect in these tasks. Finally, fine object interaction depends strongly on further factors such as grip control and tactile feedback~\cite{johansson2009coding}. Without haptic feedback, reach-to-grasp coordination in VR differs from that in the physical world~\cite{furmanek2019coordination}, which might have influenced the results in our comparison between conditions.

Beyond task-dependent effects, we also observed significant inter-individual variability. Some exhibited lower tremor-band power in VR across most or all activities, but others showed variable tremor across tasks. This variable response mirrors broader clinical realities in ET, where neither first-line pharmacotherapy \cite{deuschl2011treatment, ferreira2019mds} nor recent non-invasive neuromodulation interventions \cite{montero2025emerging} consistently suppress tremor across all patients. This variability suggests that individual characteristics can play an important role in determining the response to the altered hand visual feedback. Such heterogeneity motivates future approaches that adapt visual feedback to the characteristics and needs of individual users, rather than applying the same manipulation generally. 


Importantly, the tremor reductions observed in certain tasks in VR occurred despite patients taking equal or longer completion times in all non-time-constrained tasks compared to RW. Extended task durations and virtual interaction challenges (such as the lack of tactile feedback or unfamiliarity with the medium) could induce physical fatigue or frustration, which can increase tremor amplitude by causing emotional and physical stress \cite{paparella2026neuropsychiatric}. The fact that physical tremor expression did not escalate during these longer VR trials (and remained attenuated in specific tasks) further reinforces the hypothesis of visual feedback suppression. Future work should systematically weigh tremor expression against completion time, accuracy, and user experience.
\subsection{Limitations}
The experiment with patients involved a small sample of eight individuals with ET, which limits the statistical power and the generalizability of our findings. Furthermore, our measurements were obtained during a single experimental session, which prevents us from assessing longitudinal changes or potential adaptation effects. 

In terms of experimental design, although we attempted to reproduce the same activities in both environments, the absence of physical interaction with virtual objects, hand-tracking inaccuracies, or differences in execution speed could have influenced the measured tremor. Besides, the native hand-tracking pipeline of the Meta Quest 3 is also a proprietary system whose filtering characteristics cannot be directly controlled.

Finally, although we observed lower tremor-band power in some tasks and participants, our study does not establish a clinical or therapeutic benefit. Additional investigations with larger patient cohorts and more explicit control over the visual manipulations are necessary to determine whether such changes can translate into meaningful improvements in daily living activities.
\section{Conclusions}
\label{sec:07_Conclusions}

In this work, we have investigated how VR alters the visual representation of hand tremor and examined whether these alterations lead to changes in its physical expression. We have developed a VR tool inspired by standardized medical tremor assessments, and tested it within ecologically valid environments. Our results show that the native hand tracking of a consumer-level headset can attenuate the visual representation of high-frequency involuntary motion while allowing users to complete daily manual tasks. Our subsequent study with patients with ET revealed heterogeneous, although promising, results: several activities showed consistent reductions in measured tremor-band power across participants. Importantly, these changes depended on both the activity performed and the individual. This opens an interesting opportunity to exploit altered visual feedback as a tool for studying the interaction between vision and motor expression for tremor.


\subsection{Future work}

Our work opens several avenues for extending visual manipulation of hand motion beyond tremor attenuation. One particularly promising direction is manipulating the trajectory of goal-oriented hand movements (e.g., reaching objects) to both attenuate tremor and gradually guide patients toward the intended zone. This is especially useful given previous findings of impaired reaching adaptation in patients with ET ~\cite{blondiaux2023impairments}.

Future studies could also explore more realistic means of visual feedback, such as augmented reality, where virtual hands could overlap directly onto the user's real hands, while leaving the rest of the environment real. Additionally, implementing dynamic filtering schemes with adjustable attenuation levels would enable the study of instantaneous tremor responses during real-time parameter changes or ON/OFF toggling.

Finally, combining these techniques with more advanced wearable or haptic interfaces, such as instrumented gloves, could help investigate how altered visual, proprioceptive, and tactile feedback can jointly shape motor behavior.


\section*{Acknowledgments}
The authors wish to thank all the participants of the experiments. This work has been partially supported by the project TREMOvR, funded through the 2025 IMPULSO call by the Aragón Institute for Engineering Research (I3A); by the European Union (ERC grant number 101220555, PROXIE); by the Government of Aragon's Departamento de Educación, Ciencia y Universidades through the project “HUMAN-VR: Development of a Computational Model for Virtual Reality Perception” (PROY\_T25\_24); by European Union’s Horizon Europe Research and Innovation Program through the Marie Skłodowska-Curie under Grant 101151398; by the Government of Aragon (DGA) through the 2024 Grant Call for Doctoral research; and by MICIU/AEI and FEDER, UE under Grant PID2022-138585OA-C32 and PID2022-138585OA-C33. Views and opinions expressed are however those of the author(s) only and do not necessarily reflect those of the European Union or the European Research Council Executive Agency. Neither the European Union nor the granting authority can be held responsible for them.


\bibliographystyle{IEEEtran}
\bibliography{template}

\end{document}